\documentclass[12pt,a4paper,tightenlines,nofootinbib]{revtex4}

\usepackage{amssymb}
\usepackage{amsmath}
\usepackage{epsf}
\usepackage{psfrag}

\newcommand{\beq}{\begin{equation}}
\newcommand{\eeq}{\end{equation}}

\def\rL{\rho_\Lambda}
\def\rD{\rho_D}
\def\nablaslash{\not{\hbox{\kern-3pt $\nabla$}}}

\def\L{\Lambda}
\def\rL{\rho_\Lambda}
\def\rD{\rho_D}

\begin{document}

\author{J. Garriga$^{1,2}$ and A. Vilenkin $^2$}
\affiliation{
$^1$ Departament de F{\'\i}sica Fonamental, Universitat de Barcelona,
Mart{\'\i}\ i Franqu{\`e}s 1, 08193 Barcelona, Spain}
\affiliation{$^2$ Institute of Cosmology, Department of Physics and Astronomy,
Tufts University, Medford, MA 02155, USA}
\title{Testable anthropic predictions for dark energy}
\date{\today}

\begin{abstract}

In the context of models where the dark energy density $\rD$ is a
random variable, anthropic selection effects may explain both the
"old" cosmological constant problem and the "time coincidence". We
argue that this type of solution to both cosmological constant
problems entails a number of definite predictions, which can be
checked against upcoming observations.  In particular, in models
where the dark energy density is a discrete variable, or where it
is a continuous variable due to the potential energy of a single scalar field, the
anthropic approach predicts that the dark energy equation of state
is $p_D=-\rho_D$ with a very high accuracy. It is also predicted that the dark
energy density is greater than the currently favored value
$\Omega_D\approx 0.7$. Another prediction, which may be testable
with an improved understanding of galactic properties, is that the
conditions for civilizations to emerge arise mostly in galaxies
completing their formation at low redshift, $z\approx 1$. Finally,
there is a prediction which may not be easy to test
observationally: our part of the universe is going to recollapse
eventually. However, the simplest models predict that it will take more than a trillion years of accelerated expansion before this happens.

\end{abstract}

\maketitle

\section{Introduction}

The ``old'' cosmological constant problem - why don't we see the
large vacuum energy density $\rL$ which is expected
from particle physics? - and the ``time coincidence'' problem - why do we
live at the epoch when the dark energy component $\rD$ starts dominating? -
may find a natural explanation in models where $\rD$ is a random variable.
The idea is to introduce a dynamical dark energy
component $X$ whose contribution $\rho_X$ varies
from place to place, due to processes which occurred in the early universe.
Then
$$
\rho_D=\rho_\L +\rho_X
$$
will also vary from place to place, and the
old cosmological constant problem  takes a
different form. The question is not why $\rho_\L$ is much smaller than
$\eta^4$, where $\eta$ is some high energy physics mass scale, such as
the supersymmetry breaking scale $\eta\sim TeV$, but
why do we happen to live in a place where $\rho_\L$ is almost
exactly cancelled by $\rho_X$.
This line of enquiry is rather quantitative, since we can ask
what is the probability for us to observe certain values of
$\rho_D \sim 10^{-11} (eV)^4$,
or what is the probability for the time coincidence.

Explicit particle physics models for a variable $\rho_X$
have been reviewed in \cite{solutions}. Two examples which have
been thoroughly discussed in the literature are a four-form
field strength, which
can vary through nucleation of membranes \cite{teitelboim,alexgia},
and a scalar
field with a very low mass \cite{likely,alexgia}.
Assuming one such mechanism, and using a theory of initial conditions such as
inflation, one can calculate the "a priori" probability distribution
${\cal P}_*(\rD) d\rD$. This is defined as the fraction of co-moving
volume which
at some fiducial initial time (which we conventionally take to be the time
of recombination) had
the value of the dark energy density in the interval $d\rD$.
Inflation is also responsible for smoothing out the value of $\rD$
over comoving distances much larger than the size of our presently
observable universe.

By itself, ${\cal P}_*$ is not sufficient to calculate
probabilities for our observations. Selection effects which bias
the measurement of $\rD$ must be included, and the most important
one in this case is anthropic
\cite{carter74,carterbio,bt,Davies,Weinberg87}.
\footnote{Anthropic selection effects associated with the possible
variation of the amplitude of density fluctuations \cite{TR,mario}
and of the baryon to photon ratio \cite{TR,aguirre} have also been
discussed in the literature.} While $|\rD|$ may be very large in
most places, there is nobody there to observe such extreme values.
If $\rho_D>0$, galaxy formation stops once the dark energy becomes
dominant over the matter density. Some galaxies are seen at
redshifts of order $z\sim 5$, but not much higher, indicating
\cite{Weinberg87} that galaxies will not form in regions where
$\rho_D \gtrsim (1+z_{EG})^3 \rho_0$. Here, $\rho_0$ is the matter
density at the present time $t_0$, and $z_{EG} \approx 5$ is the
redshift at the time $t_{EG} \sim (1+z_{EG})^{-3/2} t_0$ when the
earliest galaxies formed. Also, for a negative $\rD$ the universe
recollapses on a time scale $t_{D} \sim |G\rD|^{-1/2}$, where $G$
is Newton's constant. This time should be larger than the earliest
time $t_{EI}$ which is required for intelligence to develop
\cite{bt,kalloshlinde}. Thus, observers will only exist within a
tiny ``anthropic range''
\begin{equation}
-(G t_{EI}^2)^{-1} \lesssim \rD \lesssim (G t_{EG}^2)^{-1}.
\label{range}
\end{equation}

It should be noted that, aside from the above minimal requirements,
anthropic
selection includes all other ways in which $\rD$
disfavours the existence of observers. For instance, in regions where
$\rD<0$, the matter density is larger than $|\rD|$ throughout the cosmic
evolution. If $|\rD|$ is too large, all galaxies formed in that region
will be very dense, and as a result, very inhospitable. This occurs
also  for a large $\rD>0$, since galaxies must form before $\rD$ starts
dominating. We shall come back to this issue in Sections III and V.

The selection effect can be implemented quantitatively by assuming
the mediocrity principle, according to which our civilization is
typical in the ensemble of all civilizations in the universe. The
probability to find ourselves in a region with given values of
$\rho_D$ is thus given by \cite{mediocrity}
\begin{equation}
d{\cal P}(\rD) \propto {\cal P}_*(\rD) n_{civ}(\rD) d\rD.
\label{distribution}
\end{equation}
Here, $n_{civ}(\rD)$ refers to the number of civilizations
which will ever form per unit co-moving volume in regions where the
dark energy density was equal to $\rD$ at the time of recombination.
\footnote{As we shall argue, in
models where both cosmological constant problems can be solved anthropically,
$\rD$ has not varied appreciably since the time of recombination, and therefore
it can be treated as constant in time.}

Needless to say, the determination of both
factors in the r.h.s. of Eq. (\ref{distribution}) leaves room for
some uncertainties. However, we shall argue that there are reasons to be
optimistic. If the distribution (\ref{distribution}) is
to explain both cosmological constant problems, then a number
of rather generic predictions can be made, rendering these ideas
very testable.

In the next section, we review the calculation of the prior
probability distribution ${\cal P}_*(\rD)$.  The anthropic factor
$n_{civ}(\rD)$ is discussed in section III.  In the same section, we
argue that the anthropic approach can succeed only if the conditions
for civilizations to evolve arise mostly in galaxies formed at low
redshifts, $z\sim 1$.  Anthropic predictions for the dark energy
equation of state, for the energy density $\rD$, and for the Hubble
parameter $h$ are discussed in sections IV, V.  The prediction for the
future of the universe in unveiled in section VI.  Finally, our
conclusions are briefly summarized and discussed in section VII.

\section{The prior distribution}

The first task in determining (\ref{distribution}) is to estimate
${\cal P}_*$.
The vacuum energy density is of order
$\rL \sim \eta^4\gtrsim (TeV)^4$, and therefore
$\rho_D$ must have a natural range of variation of order $\eta^4$ or larger.
Weinberg noted \cite{Weinberg87} that a function ${\cal P}_*(\rD)$ that varies smoothly
on scales $\rD \sim \eta^4$, should
behave as a constant in the utterly narrower interval (\ref{range})
- unless of course, the function would happen to have a
zero or a pole in that interval (which would be an utter
coincidence). This led him to conjecture that for values of $\rD$
in the anthropic range the prior probability would be constant,
\begin{equation}
{\cal P}_*(\rD)\approx const.
\label{conjecture}
\end{equation}
Outside of this range the form of
$\cal P_*$ is irrelevant, because the factor $n_{civ}$ vanishes.
Weinberg's conjecture is subject to verification.
As mentioned in the Introduction,
${\cal P}_*$ is calculable, provided that the dynamics of
$\rho_X$ is known, and
assuming an inflationary model which would determine its
spatial distribution at the time of recombination.
Analysis of explicit models shows that (\ref{conjecture})
is not automatically guaranteed \cite{likely}, but it does seem to be
satisfied in generic models.

There are basically two reasons \cite{likely,solutions} why
a non-flat ${\cal P}_*$ may result from the process
of randomization of $\rD$ which occurs during inflation (this randomization
is due to quantum diffusion in the case where $X$ is a scalar field,
or to nucleation of membranes in the case when $X$ is a four-form).
The first reason is the differential expansion induced by the dark
energy component. During inflation, the expansion rate is determined by
$H^2=(8\pi G/3)(V_{inf} + \rD)$.
Although $\rD$ is very small compared with the inflationary potential
$V_{inf}$, its effect may build up over time, in such a way that
more thermalized volume is generated with high values of $\rD$. In this
way, ${\cal P}_*(\rD)$ could be biased towards large values.
Let us denote by $\tau(X,H)$ the characteristic
time needed for the dynamics of $X$ to sample (at a fixed point in space)
all values of $\rD$ within the anthropic range $(\Delta \rD)_{anth}$.
The differential expansion is characterized by the parameter
\begin{equation}
q = (\Delta H) \tau = ({4\pi G/ 3}) H^{-1} (\Delta\rD)_{anth} \tau(X,H).
\label{p}
\end{equation}
If $q\gg 1$, then ${\cal P}_*$ is exponentially
steep in the range of interest. This case is ruled out by
observations, because it predicts a very large $\rD$, even
after selection effects have been factored in.
If $q\sim 1$, the distribution
${\cal P}_*$ may have a moderate dependence on $\rD$ within the
anthropic range. This dependence affects the
position of the peak of the distribution for the observed values
of $\rD$, Eq. (\ref{distribution}), and hence it affects our
predictions. While models of this sort are not ruled out,
they require a very unnatural adjustment of parameters,
since $q$ is determined by a combination of rather different
pieces of dynamics. Hence, we shall disregard this marginal
possibility as non-generic.  Finally, there is a wide class of models where
$q \ll 1$ is satisfied without any fine tuning
\cite{likely,solutions}, and hence we shall
take this to be the generic case. Numerical simulations confirm that
in this case the bias effect due to differential expansion is
insignificant \cite{VVW}.

The second reason why ${\cal P}_*$ may be non-flat is the following.
Even if the differential expansion is negligible, and the prior
distribution for $X$ is flat, this does not automatically guarantee that
the prior for $\rD$ will be
flat, unless the relation between $X$ and $\rD$ is linear in the range
of interest. Through this effect, it is possible to have a moderate
variation of ${\cal P}_*(\rD)$ within the anthropic range. But
again, this would require a contrived adjustment of parameters and we
shall dismiss this case as non-generic (see also \cite{weinbergprior} for
a discussion of this issue).

As an example, let us
consider the case where $\rho_X=V(\phi)$ is
the potential energy density of a scalar field $\phi$,
\begin{equation}
\rD= \rL + V(\phi).
\label{tr}
\end{equation}
The field must change very slowly on a cosmological
time-scale, so that its potential energy behaves as an
effective cosmological constant. This requires the slow-roll conditions
\cite{likely}
\begin{equation}
|V'|\ll 10 \rD /m_p,  \quad\quad |V''|\ll 10^2 \rD /m_p^2
\label{sr}
\end{equation}
to be satisfied up to the present time (when
$\rD \sim \rho_0$, with $\rho_0$ the present matter
density).
The constraint $q \ll 1$ on the differential expansion
yields \cite{likely}
\begin{equation}
V'^2 H^4/ GV^3 \gg 1.
\label{difexp}
\end{equation}
During inflation, the scalar field is randomized by quantum fluctuations,
and at recombination it is distributed according to the ``length'' in
field space,
\begin{equation}
{\cal P}_*(\phi) d\phi \propto d\phi.
\label{length}
\end{equation}
Therefore
 \footnote{Note that near the points where $V'(\phi)=0$, we have
$\rD \approx A + B \phi^2$ and $V'(\phi) \sim \phi \sim (\rD - A)^{1/2}$,
which is integrable. Hence, the zeroes of $V'(\phi)$ are not a concern.},
\begin{equation}
{\cal P}_*(\rD) d\rD \propto {d\rD \over |V'(\phi)|}.
\label{nonflat}
\end{equation}
Thus, the flatness of the prior depends on how much $V'$
changes in the anthropic range. As we shall see,
variations in this range may occur, but
they do not bias the probability distribution for $\rD$
in any significant way, unless we adjust some parameters specifically
for this purpose.

Consider a potential of the form
\begin{equation}
V(\phi) = {1\over 2} \mu^2 \phi^2,
\label{quadratic}
\end{equation}
where $\mu^2 \rL <0$, so that it is possible to have $|\rD|$
very small even if $|\rL|$ is large.
Eqs. (\ref{sr}) lead to the condition \cite{likely}
\begin{equation}
|\mu| \ll 10^{-120} m_p^3 |\rL|^{-1/2}.
\label{strongconst}
\end{equation}
Such a small mass parameter may seem unrealistic, but it can naturally
arise, for instance, in a low energy effective theory with a
suitable discrete symmetry \cite{alexgia} (for other proposals, see
\cite{solutions,weinbergprior,donoghue} and references therein).
Note that (\ref{strongconst}) does not correspond to a fine tuning,
but just to a strong suppression.
The condition (\ref{difexp})
translates into
\begin{equation}
|\mu| \gg H_0^3 H^{-2} \sim 10^{-169} m_p,
\label{upperconst}
\end{equation}
where $H_0$ is the present Hubble rate, and
in the last step we have used $H\sim 10^{-7} m_p$,
corresponding to a GUT scale of inflation. The conditions
(\ref{strongconst}) and (\ref{upperconst}) leave very many
orders of magnitude available for the parameter $\mu$, and so
fine tuning is not necessary.
From (\ref{tr}),
\begin{equation}
\rD=\kappa (\phi-\phi_0) +{\mu^2\over 2}(\phi-\phi_0)^2,
\label{expansion}
\end{equation}
where $\phi_0^2=-2\rL/\mu^2$ and $\kappa =\mu^2 \phi_0$.
We are interested in the vicinity of $\rD=0$, where it is
easy to show from (\ref{nonflat}) that \cite{likely}
\begin{equation}
{\cal P}_*(\rD) d\rD \propto [1+{\cal O}(\rD/\rL)] d\rD \approx d\rD.
\label{veryflat}
\end{equation}
Since $\rD \ll \rL$ in the anthropic range, the distribution is indeed
flat to a very good accuracy.

For contrast, we may consider the ``washboard'' potential
\begin{equation}
\rD= \rL+ \kappa \phi + M^4 \sin(\phi/\eta),
\label{washboard}
\end{equation}
where $\kappa$ was given above and $M$ and $\eta$ are different mass scales.
Let us assume that
\footnote{If $M^4 \gg H_0^2 m_p \eta$, the slow roll condition is
not satisfied today and the field $\phi$ will be in any one
of the local minima of the washboard. With some generic requirements
on the inflationary parameters, the minima will have equal a priori
probability within the anthropic range \cite{solutions}.}
\begin{equation}
M^4 \ll H_0^2 m_p \eta \sim (\eta/m_p) \rho_0.
\label{m4}
\end{equation}
Then the field will typically be found away from the
local minima, with a probability distribution
\begin{equation}
{\cal P}_*(\rD) d\rD = {d\rD \over |\kappa+ (M^4/\eta) \cos(\phi/\eta)|^{-1}}.
\label{notsoflat}
\end{equation}
Both $\kappa$ and $M^4/\eta$ should be much smaller
than $H_0^2 m_p$ in order to satisfy the slow roll condition.
In the case $\kappa \gg M^4/\eta$, the distribution
(\ref{notsoflat}) is still flat, as in (\ref{conjecture}).
In the opposite case, where $M^4/\eta \gg \kappa$, the a priori
distribution can have a sizable variation within the
anthropically allowed range.
If $\eta \ll m_p$, this range
is very wide in the field space, $\delta\phi \gtrsim \rho_0/\kappa \gg m_p$.
This means that the oscillations in ${\cal P}_*$ will
average out on scales much smaller than the anthropic range, and
effectively we recover (\ref{conjecture}). Clearly, the only
way to avoid this averaging effect is if $\eta\gtrsim m_p$, and
\begin{equation}
M^4 \sim (\Delta\rD)_{anth}.
\label{m4a}
\end{equation}
The last equation is to ensure that a significant range of values of $\phi/\eta$ is
sampled in the anthropic range $(\Delta\rD)_{anth}\lesssim 10^3 \rho_0$, so that changes
in the slope of the potential are appreciable. Otherwise the distribution for ${\cal P}_*$
will be almost flat. Thus, aside from the fact that the washboard potential is already
a somewhat contrived example \cite{weinbergprior},
Eq. (\ref{m4a}) implies an otherwise unnecessary adjustment
of the parameter $M$.

In what follows, we shall only consider models where there
is no such ad-hoc adjustment. In this sense,
our predictions may not be completely inescapable, but they
can be considered generic. The situation can be compared with the
predictions of inflation that the density parameter is $\Omega=1$ and
the spectrum of density perturbations is nearly flat. It is certainly
possible, in the context of inflation, to have an open universe with
$\Omega<1$, or to
have a markedly non-flat spectrum of density perturbations. But to
 achieve this, additional parameters must be introduced and adjusted to
the desired outcome.

\section{The anthropic factor}

We now consider the effect of the anthropic factor $n_{civ}$ in
Eq. (\ref{distribution}).  The physical situation is rather different
for positive and negative $\rD$, so we consider these two cases
separately.


For positive $\rD$, the main change introduced by $n_{civ}$ is that
the time of earliest galaxy formation $t_{EG}$ in the anthropic range
(\ref{range}) is effectively replaced by
the time at which the bulk of galaxy formation occurs.
This is because a few early birds will not make a
difference once we apply the principle of mediocrity.
More precisely, we should
take into consideration that the morphology of some galaxies could
make them less suitable for the development of civilizations, and therefore
\begin{equation}
n_{civ}(\rD) = \int\ d\alpha\ n(\alpha,\rD)\ N_{civ}(\alpha).
\label{nciv}
\end{equation}
Here, $\alpha$ denotes the set of parameters characterizing the type of galaxy
(e.g. its size, density, etc.), $n(\alpha,\rD)$ is the number density of
such galaxies that form per co-moving volume
in regions characterized by $\rD$, and $N_{civ}(\alpha)$
is the number of civilizations per galaxy  of type $\alpha$.
Suppose that the above integral receives
a dominant contribution from galaxies of type $\alpha_G$. Then
\beq
n_{civ}(\rD)\propto n(\alpha_G,\rD),
\label{ncivrD}
\eeq
and  the
relevant time for anthropic considerations is the time at which this
type of galaxies form, which we shall denote by $t_G$. With the
assumption of a flat prior ${\cal P}_*$, it was shown in
\cite{mario,Bludman} that the most probable value for a positive $\rD$
is the one characterized by
\begin{equation}
t_D \sim t_G.
\label{tcoincidence}
\end{equation}
This fact was used in order to explain the observed time coincidence
\begin{equation}
t_D \sim t_0.
\label{tdt0}
\end{equation}
The last relation follows from (\ref{tcoincidence}), assuming that
stars and civilizations develop on a timescale not much greater than $t_G$,
and therefore $t_G$ is comparable to $t_0$, defined as the time
when most civilizations make their first determination of $\rD$.

Connected with the above discussion, there is a prediction of the
anthropic approach, which can be checked by a combination of
observations and theoretical analysis.  In a not so distant future,
our understanding of galactic evolution and morphology may improve to
the point where we can tell with some confidence which galaxies are
suitable for sustaining planetary systems similar to our own, where
civilizations can develop. The anthropic approach to the cosmological
constant problems (CCPs) predicts that the conditions for civilizations to
emerge will be found mostly in galaxies that formed (or completed
their formation) at a low redshift, $z\sim 1$.

In the standard cold dark matter cosmology, galaxy formation is a
hierarchical process, with smaller objects merging to form more and
more massive ones.  We know from observations that some galaxies
existed already at $z = 5$, and the theory predicts that some dwarf
galaxies and dense central parts of giant galaxies could form as
early as $z=10$ or even $20$.
The fraction of matter bound in giant galaxies ($M\sim 10^{12}M_\odot$)
at $z=1$ ($\sim 20\%$) is somewhat less than that in objects of mass
$\sim 10^{9}M_\odot$ at $z=3$, or in objects of mass $\sim 10^7
M_\odot$ at $z=5$ \cite{MoWhite}.  If civilizations were as likely to
form in early galaxies as in late ones, then
Eq. (\ref{tcoincidence}) would indicate that, for a typical
observer, the cosmological constant should start dominating at a
redshift $z_G\gtrsim 5$. The
corresponding dark energy density,
\beq
\rD \sim (1+z_G)^3 \rho_0,
\label{rDzG}
\eeq
would be far greater than observed. Clearly, the
agreement becomes much better if we assume that the conditions for
civilizations to emerge arise mainly in the types of galaxies which form
at lower redshifts, $z_G \sim 1$.

We now point to some directions along which the
choice of $z_G\sim 1$ may be justified.  One problem with dwarf
galaxies is that if the mass of a galaxy is too small, then it cannot
retain the heavy elements dispersed in supernova explosions.
Numerical simulations suggest that the fraction of heavy elements
retained is $\sim 30\%$ for a $10^9 M_\odot$ galaxy and is negligible
for much smaller galaxies \cite{MacLow}.  The heavy elements are
necessary for the formation of planets and of observers, and thus one
has to require that the structure formation hierarchy should evolve up
to mass scales $\sim 10^{9}M_\odot$ or higher prior to the dark
energy domination.  This gives the condition $z_G\lesssim 3$, but
falls short of explaining $z_G\sim 1$.

Another point to note is that smaller galaxies, formed at earlier
times, have a higher density of matter.  This may increase the danger
of nearby supernova explosions and the rate of near encounters with
stars, large molecular clouds, or dark matter clumps.  Gravitational
perturbations of planetary systems in such encounters could send a
rain of comets from the Oort-type cloud towards the inner planets,
causing mass extinctions\footnote{The cross-section for disruption of
planetary orbits is much smaller, and it would take a rather
substantial increase of the density for this process to become
statistically important.  A.V. is grateful to David Spergel for a
discussion of this issue.}.

Our own Galaxy has definitely passed the test for the evolution of
intelligence, and the principle of mediocrity suggests that most
observers may live in galaxies of this type.  Our Milky Way is a giant
spiral galaxy.  The dense central parts of such galaxies were formed
at a high redshift $z\gtrsim 5$, but their discs were assembled at
$z\sim 1$ or later \cite{Abraham}.  Our Sun is located in the disc, at
a distance $\sim 8.5$ kpc from the Galactic center\footnote{It has been noted
\cite{Leitch} that this distance is close to the corotation radius,
where the orbital velocity of the stars coincides with the rotational
velocity of the spiral pattern.  In other words, the motion of the Sun
relative to the spiral arms is rather slow, and as a result, the
periods between spiral arm crossings are rather long ($\sim 10^8$
yrs).  Spiral arms are the primary sites of supernova explosions.
They are also rich in giant molecular clouds, and are therefore very
hazardous to life.  It has been argued in \cite{Leitch} that spiral
arm crossings are responsible for the major mass extinctions observed
in the fossil record.  Then one expects that habitable planetary
systems are to be found mainly in the vicinity of the corotation
radius, since mass extinctions at a rate much greater than once in
$10^8$ yrs may be too frequent for intelligent life to evolve.  (Note
that it took us $6.5\times 10^7$ yrs to evolve since the last great
extinction.)}.  If this situation is typical, then the relevant epoch to
use in Eq. (\ref{rDzG}) is the epoch $z_G\sim 1$ associated with the
formation of discs of giant galaxies.

The above remarks may or may not be on the right track, but we
emphasize once again that if CCPs have an
anthropic resolution, then, for one reason or another, the evolution
of intelligent life should require conditions which are found mainly
in giant galaxies, which completed their formation at
$z_G\sim 1$.

In order to estimate $n(\alpha_G,\rD)$ in Eq.~(\ref{ncivrD}), we shall need a
simple quantitative criterion to specify the relevant type of
galaxies.  The most important parameter characterizing a galaxy is its
mass $M$.  For the Milky Way it is $M_{MW}\sim 10^{12}~M_\odot$
\cite{Wilkinson}, and the above discussion suggests that we identify
the relevant galaxies with gravitationally bound halos of this mass.
(Note that this is also the typical mass of $L_*$ galaxies, which
contain most of the luminous stars in the Universe.)  It should be
recognized, however, that the choice of this characteristic mass scale is
somewhat uncertain, so we shall illustrate how our results are
affected by choosing a larger or a smaller mass.

Our Galaxy is a member of the Local Group cluster, whose mass has been
estimated as \cite{Peebles} $M_{LG}\sim 4\times 10^{12}~M_\odot$.  It
is conceivable that the gas captured in this cluster is later accreted
onto the member galaxies and thus affects the properties of their
discs.  There seems to be no justification to consider larger mass
objects, and we shall regard $M_{LG}$ as an upper bound on the
potentially relevant mass scales.  On the lower mass end, we shall use
$M\sim 10^{11}~M_\odot$, which is roughly the mass of the bright part
of our Galaxy, up to $\sim 10$ kpc from the center.  (We note that
$M_{MW}$ is probably a more reasonable choice, because the properties
of the disc depend on the total mass of the halo \cite{White}.)


We now consider negative $\rD$.  The scale factor of a universe filled
with nonrelativistic matter and dark energy with $\rD<0$ is given by
\beq
a(t)=\sin^{2/3}\left({t\over{t_D}}\right),
\eeq
where $t_D\equiv (1/6\pi G|\rD|)^{1/2}$.
The matter density
$\rho_M$ initially decreases while the universe expands, but at $t=\pi
t_D/2$, when it
reaches the value $\rho_M=-\rho_D$, the universe stops its expansion
and starts recontraction.   The matter density grows in the
contracting phase, and thus $\rho_M \geq |\rD|$ throughout the
evolution.  The structure formation in a universe with a negative
$\rD$ proceeds as usual until $t\sim t_D$, but then the
growth of density perturbations accelerates during the contraction, so
that all overdensities collapse to form bound objects prior to the big
crunch.  For $t_D\gtrsim t_0$, giant galaxies will form at about the
same time as they did in our part of the universe and will have
similar properties (with a possible caveat indicated below).  However,
for $t_D \ll t_0$ halos of the galactic size will be forced to
collapse at a much earlier time $t\sim t_D$, and their density will
therefore be much higher than that of our Galaxy.  This would probably
make such halos unsuitable for life.

These considerations suggest that the anthropic factor effectively
constrains $t_D$ to be in the range
\beq
t_D\gtrsim t_0
\label{symmetric}
\eeq
for both positive and negative $\rD$.  There is, however, an additional
factor that could make negative $\rD$ less probable.  For $\rD>0$,
structure formation effectively stops at $t>t_D$, and the existing
structures evolve more or less in isolation.  This may account for the fact
that discs of giant galaxies take their grand-design
spiral form only relatively
late, at $z\sim 0.3$.  The discs are already in place at $z\sim 1$, but
they have a very unsettled, irregular appearance \cite{Abraham}.
On the other hand, for $\rD<0$ the clustering hierarchy only speeds up at
$t>t_D$, and quiescent discs which may be necessary for the evolution
of fragile creatures like ourselves may never be formed.

Another factor to consider is the characteristic time $t_I$ needed for
intelligence to develop.  For positive $\rD$, this factor is
unimportant, since the time after the dark energy domination is
practically unlimited, but for negative $\rD$ the available time is
bounded by $t< \pi t_D$, and the effect of $t_I$
requires a closer examination.

We first note that $t_I \ll t_0$ is unlikely, since then it is not
clear why it took so long for intelligence to develop on Earth.  (The
total time of biological evolution, from the origin of life on Earth
till present, is estimated at $\sim 3.5\times 10^{9}$ yrs.)
For $t_I \gg t_0$, we note that the main sequence lifetime of stars
believed to be suitable to harbor life is $t_\star \sim (5 - 20)\times
10^9 {\rm yrs} \sim t_0$ (see \cite{mario} for a discussion of this
point).  If $t_I\gg t_0\sim t_\star$, most of these stars will explode
as red giants before intelligence has a chance to develop.  Carter
\cite{carterbio} has argued that this is the most likely
scenario\footnote{The coincidence $t_I\sim t_\star$ is unlikely, since
the evolution of life and evolution of stars are governed by
completely different processes.}.  In this case, the number $N_{civ}$
is suppressed by a factor $\sim{\rm min}\{ t_\star,t_D\}/t_I \sim
t_0/t_I$, where we have used (\ref{symmetric}) in the last step.  For
positive $\rD$, the suppression is by a factor $\sim t_\star/t_I$,
which is of the same order of magnitude.

We conclude that the precise value of $t_I$ has little effect on the
relative probability of positive and negative $\rD$.  If the
accelerated clustering hierarchy is detrimental for life, then the
probability for negative $\rD$ is suppressed; otherwise the two signs
of $\rD$ are equally likely.  In either case, we should not be
surprised that $\rD$ is positive in our part of the universe.  In the
following sections we shall focus on the positive values of $\rD$.

\section{Prediction for the equation of state}

 A rather generic prediction of models where both CCP's are solved anthropically
is that the equation of state of dark energy is given by $p_D = w\rD$, with
\begin{equation}
w = -1 \pm 10^{-5}.
\label{eqofstate}
\end{equation}
The error bars correspond to the precision to which the
observable universe can be aproximated by a homogeneous and isotropic model.
In models where $\rho_X$ is
the energy density of a four-form field, this equation of state
is guaranteed by the fact that the four-form energy density is a constant
and can only change by the  nucleation of branes (other than that, it
behaves exactly like an additional cosmological constant). If $\rho_X$ is a
generic scalar field potential, the slow roll conditions
(\ref{sr}) are likely to be satisfied by excess, by many orders
of magnitude, rather than marginally. For instance, for the quadratic
potential (\ref{quadratic}), these conditions imply the constraint (\ref{strongconst}).
It would be contrived to
arrange for the condition to be satisfied marginally, since the whole point
of the present approach is to have $\rL$ cancelled regardless of its
precise value (which is not known to us even by order of magnitude).
If the slow roll conditions are satisfied by excess by just more
than three orders of magnitude, then the kinetic energy of the scalar field
will be less than its potential energy by more than six orders of magnitude,
and Eq. (\ref{eqofstate}) follows.

There are certainly models for dark energy, some of them with
anthropic input, were (\ref{eqofstate}) is not satisfied. For
instance, Kallosh and Linde \cite{kalloshlinde} recently considered a
supergravity model where the time coincidence problem is solved
anthropically, and where (\ref{eqofstate}) does not hold. However,
their model does not solve the old CCP, since it is assumed that the
cosmological constant vanishes in the observable matter sector due to
some unspecified mechanism. Likewise, (\ref{eqofstate}) does not hold
in the usual quintessence models \cite{quintessence}, which have no
anthropic input at all, but which do not address the CCP's
\cite{solutions,weinbergprior}, or in models of k-essence
\cite{kessence}, where only the time coincidence is partially
addressed.

A possibility worth discussing is the case of models where
the slow roll parameters are themselves random fields.
Consider, for instance, the following model:
\begin{equation}
\rD = \rL + \mu^2(\psi)\phi^2.
\label{kit}
\end{equation}
If the probability distribution for the new scalar field $\psi$
were such that all values of $\mu^2$ are equiprobable, then one might imagine that the order of magnitude of  $\mu^2$ would be such that the slow roll conditions would be marginally satisfied. \footnote{The new field $\psi$ must also be a light field and hence its prior distribution is in principle calculable. We can actually 
consider a more general form of the potential for $\psi$ and $\phi$,
\begin{equation}
\rD = \rL + V(\psi,\phi).
\end{equation}
Around any point $(\psi_0,\phi_0)$ on the curve $\gamma$ defined by
$V(\psi,\phi)=-\rL$, the potential can be
approximated by a linear function of the fields.
Moreover, we can always rotate coordinates in field space so that $\psi$ is
directed along $\gamma$, and $\phi$ is orthogonal
to it,
$$
\rD \approx V_\phi(\psi_0,\phi_0) (\phi-\phi_0).
$$
Here $V_\phi$ is the gradient of the potential at that point.
During inflation, both fields $\psi$ and $\phi$
are randomized by quantum fluctuations. Hence, the prior probability
distribution is given by the area in field space
${\cal P}_*d\psi d\phi \propto d\psi d\phi$, which leads to
\begin{equation}
{\cal P}_*(\rho_D) d\rD =
\left[\int_\gamma {d\psi \over |V_\phi|}\right] d\rho_D.
\label{integamma}
\end{equation}
Along the curve $\gamma$, the values of $V_\phi$ that will carry
more weight per unit distance along the curve are those for which the slope is smaller. In the published version of this paper, it was (incorrectly) concluded from this observation that "given a model where the slope of the potential is variable, smaller values of the slope are preferred a priori, and
there is no reason to expect that the slow roll conditions should be
satisfied only marginally." However, this is not necessarily correct in general,
since the slope could be large for a very large range of $\psi$, and large slopes may end up dominating the integral in Eq. (\ref{integamma}). A detailed analysis of this point has been given in \cite{GLV03}, where it is shown that there is a wide class of two-field models for which a small slope is favoured a priori (and for which the conclusions of Section IV do indeed apply). However, there is an equally broad class of models for which a large slope is favoured a priori. In that case, the posterior probability distribution for the slope is such that the slow roll condition is satisfied only marginally, and some departure from the vacuum equation of state may be expected (see \cite{kklls,dt}).}

\section{Predictions for $\Omega_D$ and $h$.}

Currently favoured values for the dark energy density and for the
Hubble parameter are $\Omega_D\approx .7$ and $h\approx .7$
\cite{sn,hst}, both with error bars of the order of $10\%$.
While observations are not very
accurate, we would like to challenge the status quo and boldly
use the anthropic
approach to the CCP's to make predictions for these two parameters. As
we shall see, this approach predicts that $\Omega_D$ is likely to be
somewhat higher, and that $h$ is likely to be smaller than those currently
favoured values.

The basic reason why we expect $\Omega_D$ to be larger is the
following \cite{mediocrity,efstathiou}.  The growth of density
fluctuations in a universe with a positive cosmological constant
effectively stops at the redshift $z_D$ when the cosmological constant
starts dominating. This is given by $(1+z_D) \sim
(\Omega_D/\Omega_M)^{1/3}$, where $\Omega_M=1-\Omega_D$ is the matter
density parameter.  According to (\ref{tcoincidence}), we expect
$z_D\sim z_G$, where $z_G$ is the epoch when the relevant galaxies
were formed.  With $z_G\sim 1$,
this corresponds to $(\Omega_D/\Omega_M) \sim 8$, which in turn
implies $\Omega_D \sim .9$.
(For $z_G>1$, we would obtain an even higher value for $\rD$.)
This prediction
can be made more quantitative \cite{MSW} by using the distribution
(\ref{distribution}). As we shall see, the precise predictions depend
not only on $\Omega_D$ but also on $h$.

Throughout this section, we shall assume that $\rD>0$ as part of our
prior. In a universe filled with pressureless matter
and with a dark energy component $\rD>0$, the scale factor behaves as
$$
a(t) = \sinh^{2/3}\left({t\over t_D}\right),
$$
where $t_D\equiv ({1/6\pi G \rD})^{1/2}$.
A primordial overdensity will eventually collapse, provided that its value
at the time
of recombination is larger than a certain value $\delta_c^{rec}$. In the
spherical collapse model, this is estimated as
$\delta_c^{rec}(\rD)=1.13 x_{rec}^{1/3}$, where
$x_{rec} = x(t_{rec})$ \cite{masha}.
Here, we have introduced the variable
\begin{equation}
x(t) \equiv {\Omega_D(t) \over \Omega_M(t)}= \sinh^2\left({t\over t_D}\right).
\label{xt}
\end{equation}
The number of galaxies $n(M,\rD)$ of mass $M$ that will form per
unit comoving volume in a region characterized by the value $\rD$
of the dark energy density, is proportional to the fraction of
matter that eventually clusters into this type of galaxies. In the
Press-Schechter approximation \cite{PS,masha}, this is given by
\begin{equation}
n_{civ}(\rD) \propto n(M,\rD) \propto {\rm
erfc}\left({\delta_c^{rec}(\rD)\over \sqrt{2} \sigma_{rec}(M)}
\right). \label{ngal}
\end{equation}
Here, erfc is the complementary error function, and
$\sigma_{rec}(M)$ is the dispersion in the density contrast at the
time of recombination $t_{rec}$. As argued in the preceeding
section, we shall assume that most civilizations are formed in
galaxies characterized by a mass $M \sim M_{MW} \approx 10^{12}
M_\odot$ (although we shall also consider slightly larger and
smaller masses).

The factor $n_{civ}$ depends on the parameter $\sigma_{rec}$, which in
turn depends on the amplitude of density perturbations
generated during inflation.
The value of $\sigma_{rec}$ can be inferred from the normalization of
CMB anisotropies, but for this task, both the present value of $\Omega_D$
and the value of the Hubble parameter $h$ would be needed. Since these
are the parameters we wish to make predictions about, it would be
somewhat contrived to use them at this point to make an inference
about $\sigma_{rec}$.

Another factor to consider is that $\sigma_{rec}$ may
be different in distant regions of the universe (where, as a consequence,
galaxies would form earlier or later).
In models where the inflaton field has only
one component, the value of $\sigma_{rec}$ is the same in all regions
of the universe. However, if the inflaton
field has more than one component, the amplitude of density
perturbations depends on the path
followed by the inflaton on its way to the minimum of the potential.
In such models, it is possible for $\sigma_{rec}$ to vary over distances
much larger than the presently observable universe.

\begin{figure}[t]
\centering \hspace*{-4mm}
\leavevmode\epsfysize=10 cm \epsfbox{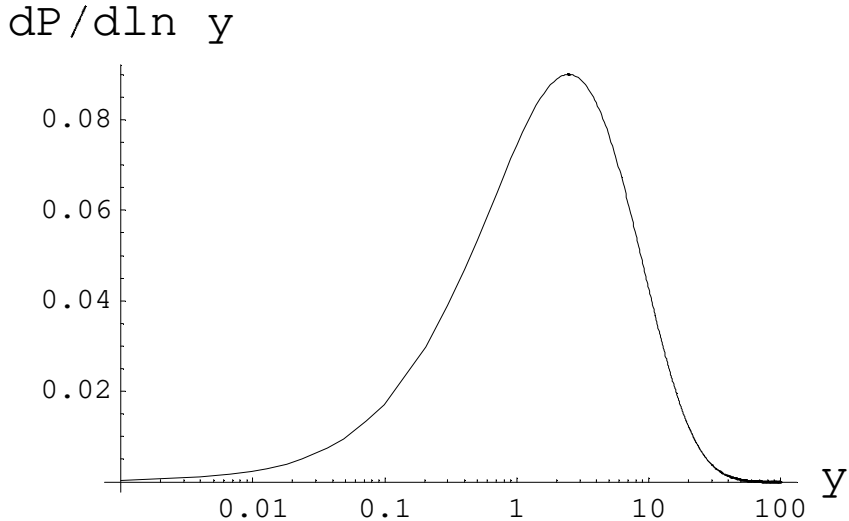}\\[3mm]
\caption[fig1]{\label{fig1} The distribution (\ref{disy}).}
\end{figure}

To make our discussion sufficiently general, we shall consider that
$\sigma_{rec}$ is itself a random variable with unspecified prior.
This prior may be determined by processes occurring during inflation,
or it may just reflect our ignorance of the actual value of the
fixed parameter $\sigma_{rec}$.
Then, Eq. (\ref{distribution}) is generalized to
\begin{equation}
d{\cal P}(\rD,\sigma_{rec}) \approx n_{civ}
{\cal P}_*(\rD,\sigma_{rec}) d\rD d\sigma_{rec}.
\label{newdistribution}
\end{equation}
In this context, the generic expectation
that the prior does not depend on $\rD$ in the anthropic range
[see Eq. (\ref{conjecture})],
translates into
$$
{\cal P}_*(\rD, \sigma_{rec}) \approx {\cal P}_*(\sigma_{rec}).
$$
Substituting (\ref{ngal}) into (\ref{newdistribution}),
we have
\begin{equation}
d{\cal P}(\rD,\sigma_{rec}) \propto
{\rm erfc}\left({.80 x_{rec}^{1/3} \over \sigma_{rec}}\right)
{\cal P}_*(\sigma_{rec})\ dx_{rec} d\sigma_{rec},
\label{disxrec}
\end{equation}
where we have used that $\Omega_M(t_{rec}) \approx 1$ in all
regions of interest, so that $d\rD \propto dx_{rec}$.
Introducing $y=x_{rec}\sigma_{rec}^{-3}$, the change of variables
$(x_{rec},\sigma_{rec})\to (y,\sigma_{rec})$ produces a Jacobian
proportional to $\sigma_{rec}^3$, and we have
\footnote{The appearance of the factor $\sigma_{rec}^3$ in the posterior distribution was noted in Ref. \cite{mario}. This factor implies that the prior $P_*(\sigma_{rec})$ should decay faster than $\sigma_{rec}^{-3}$ at large values of the density contrast. Otherwise, the posterior distribution is not normalizable (and we should expect both a large value of the effective vacuum energy and of the linear density contrast, in contradiction with observations). In the published version of the present paper this factor was omitted, due to an incorrect normalization of Eq. (\ref{disxrec}). We thank Takahiro Tanaka for drawing our attention to this point.}
$d{\cal P}(y,\sigma_{rec})\approx f(y) \sigma_{rec}^3 P_*(\sigma_{rec}) dy d\sigma_{rec}$,
where $f(y)$ does not depend on $\sigma_{rec}$.
Integrating over $\sigma_{rec}$ leads to the normalized
distribution
\begin{equation}
d{\cal P}(y) = (.80)^3 \pi^{-1/2} {\rm erfc}(.80 y^{1/3})\ y\ d\ln y,
\label{disy}
\end{equation}
which is uncorrelated with $\sigma_{rec}$.

The variable $y$ can be expressed in terms of observable
quantities, as we shall see below, and from (\ref{disy}) we should expect $y \sim 1$ by
order of magnitude (See Fig. 1). More precisely, we expect $y>.79$
with probability
\begin{equation}
P(y>.79) = .68 \quad \quad (1\sigma\ {c.l.}),
\label{1sigma}
\end{equation}
and $y>.07$ with probability
\begin{equation}
P(y>.07) = .95 \quad \quad (2\sigma\ {c.l.}).
\label{2sigma}
\end{equation}
We shall denote these two equations as the $1\sigma$ and $2\sigma$
confidence level predictions for $y$. Let us now show how these
translate into confidence level curves for the expected values of
the parameters $\Omega_D$ and $h$.
Here, and in what follows, $\Omega_D$ will denote the {\em present} value of
the dark energy density parameter in our observable universe.

Let us first express the ``observed'' value of $y$, which we shall denote
as $y_0$, in terms of $\Omega_D$ and $h$.
The density contrast at present is given by
$\sigma_0 = G(x_0,x_{rec}) \sigma_{rec}$, where, assuming $z_{rec}\gg 1$,
the growth factor is given by \cite{MSW}
$G(x_0,x_{rec}) = x_{rec}^{-1/3} F(\Omega_D)$,
with
\begin{equation}
F(\Omega_D) = {5 \over 6} \Omega_D^{-1/2} \int_0^{\Omega_D/(1-\Omega_D)}
{dw \over w^{1/6}(1+w)^{3/2}}.
\label{F}
\end{equation}
Therefore,
\begin{equation}
y_0=\left[{F(\Omega_D)\over \sigma_0}\right]^3.
\label{y0}
\end{equation}
The linearized density contrast at present $\sigma_0$ can be inferred from
measurements of CMB temperature anisotropies, as described e.g. in
\cite{ll,MSW}. Since the spectrum is expressed as a function of
wavelength, the mass scale has to be converted into a
length-scale.  A halo of mass $M$ corresponds to a co-moving
radius $R(M)=(3M/4\pi \rho_0)^{1/3}$. The mean matter density of
the universe is given by $\rho_0=1.88 \times 10^{-29} \Omega_M
h^2\ g/cm^3$, which leads to
$$
R(M)=.95 h^{-2/3}\Omega_M^{-1/3}\left({M\over
10^{12}M_{\odot}}\right)^{1/3} {\rm Mpc}.
$$
Assuming an adiabatic primordial spectrum of scalar density
perturbations, characterized by a spectral index $n$, we have
\begin{equation}
\sigma_{0}(R) = (c_{100}\Gamma)^{(n+3)/2}\delta_H K^{1/2}(R).
\label{sigma0}
\end{equation}
Here, $c_{100}= 2.9979$ is the speed of light in units of $100 km\ s^{-1}$
and
$$
\Gamma=\Omega_M h \exp[-\Omega_b(1+ \sqrt{2h}\Omega^{-1}_M)]
$$
is the so-called shape parameter, with $\Omega_b$ the density parameter in
baryons. For numerical estimates, we shall take $\Omega_b h^2 \approx .02$.
The dimensionless amplitude of cosmological perturbations inferred
from  the COBE DMR experiment is given
by \cite{bw,ll}
\begin{equation}
\delta_H= 1.91 \times 10^{-5}
{\exp[1.01(1-n)]\over \sqrt{1+r(.75-.13\Omega_D^2)}}
\Omega_M^{-.80-.05 \ln \Omega_M}[1-.18(1-n)\Omega_D-.03 r \Omega_D].
\label{deltah}
\end{equation}
The parameter
$r$ denotes the ratio of tensor to scalar amplitudes.
Note that the effect of tensors is to make $\delta_H$ a bit smaller
(although not very significantly). Finally,
$$
K(R)= \int_0^\infty q^{(n+2)}  T^2(q) W^2(q\ \Gamma\ h\ R\ Mpc^{-1}) dq,
$$
where the transfer function is given by
$
T(q) = (2.34 q)^{-1}\ln(1+2.34 q)
[1+3.89 q+(16.1 q)^2+(5.46 q)^3+(6.71q)^4]^{-1/4}
$
and the window function is given by $W(u) = 3 u^{-3}(\sin u - u \cos u)$.

Substituting (\ref{sigma0}) in (\ref{y0}), and using
$\Omega_D+\Omega_M=1$, we obtain the function $y_0 = y_0
(\Omega_D,h)$. Contour lines of this function, corresponding to
the $1\sigma$ and $2\sigma$ predictions represented by Eqs.
(\ref{1sigma}-\ref{2sigma}), are plotted in Fig. 2, assuming that
the dominant contribution to $n_{civ}$ is in galaxies of mass
$M=M_{MW}=10^{12}M_\odot$ (thick solid lines). We also consider
the predictions for different choices of the mass, as discussed in
Section III. The short dashed curves correspond to the mass of the
local group $M_{LG}=4\times 10^{12}M_\odot$, and the long dashed
curves correspond to the mass of the bright inner part of our
galaxy $M=10^{11}M_\odot$. The effect of a tilt in the spectral
index is plotted in Fig. 3. Both of these figures ignore the
effect of tensor modes in the normalization (\ref{deltah}). Tensor
modes tend to lower the value of $\delta_H$, and hence they tend
to make the bounds somewhat less stringent. The effect, however,
is not dramatic. Even for $r$ as large as $.5$, the effect on the
curves is comparable to the effect of lowering the spectral index
by $.05$.

\begin{figure}[t]
\psfrag{h}[][r]{$h$}\psfrag{W}[][r]{$\Omega_D$}\psfrag{s}{$\sigma$}
\centering \hspace*{-4mm}
\leavevmode\epsfysize=10 cm \epsfbox{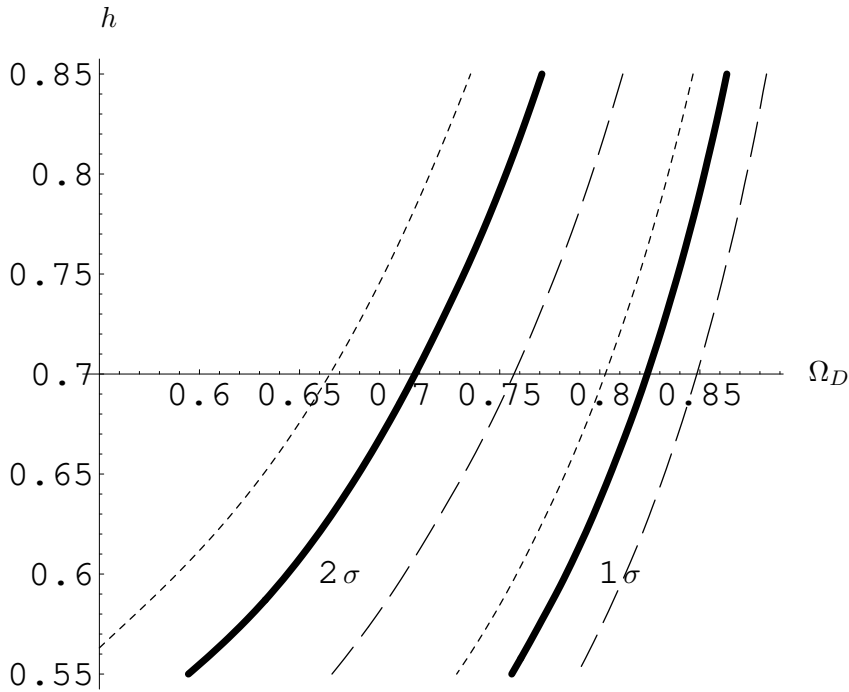}\\[3mm]
\caption[fig2]{\label{fig2} Contours of the function
$y_0(\Omega_D, h)$ given in (\ref{y0}), corresponding to the
$1\sigma$ (lower curves) and $2\sigma$ (upper curves) predictions
represented by Eqs. (\ref{1sigma}-\ref{2sigma}).
The excluded region lies to the left of the curves.
The thick solid
lines assume that the dominant contribution to $n_{civ}$ is in
galaxies of mass $M=M_{MW}=10^{12}M_\odot$. For comparison, we
show the predictions for different choices of the mass. The short
dashed curves correspond to the mass of the local group
$M_{LG}=4\times 10^{12}M_\odot$, and the long dashed curves
correspond to the mass of the bright inner part of our galaxy
$M=10^{11}M_\odot$. A scale invariant spectrum of density
perturbations is assumed.}
\end{figure}

\begin{figure}[t]
\psfrag{h}[][r]{$h$}\psfrag{W}[][r]{$\Omega_D$}\psfrag{s}{$\sigma$}
\centering \hspace*{-4mm}
\leavevmode\epsfysize=10 cm \epsfbox{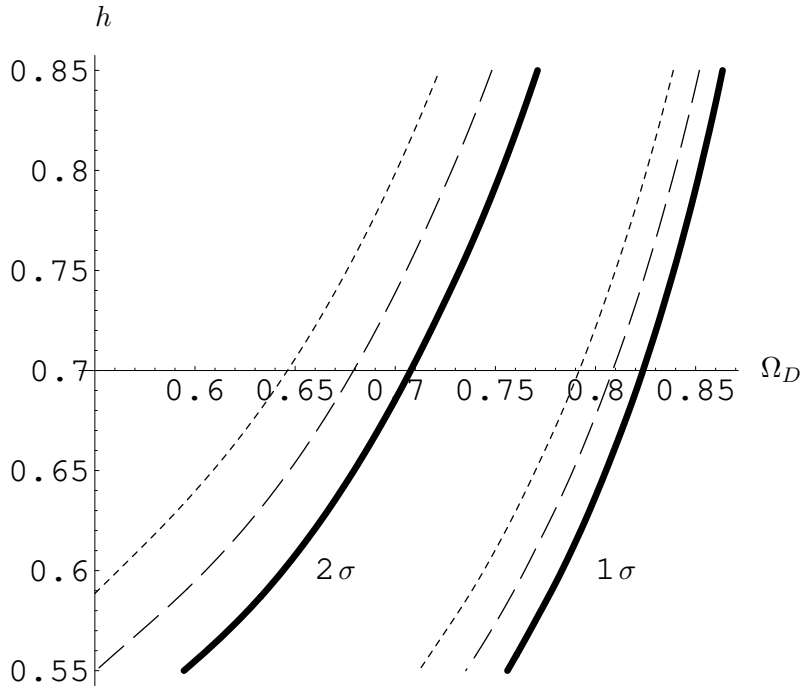}\\[3mm]
\caption[fig3]{\label{fig3} Effect of a tilt in the spectral index
of density perturbations. As in Fig. 2, the thick solid lines
correspond to a scale invariant spectrum $n=1$, and a mass
$M=M_{MW}=10^{12}M_\odot$. The long dashed line and the short
dashed lines correspond to tilted spectra, with $n=.95$ and $n=.9$
respectively.}
\end{figure}

Expressions similar to Eqs.~(\ref{disy})-(\ref{y0}) were already contained
in the exhaustive analysis of the problem given by Martel, Shapiro and
Weinberg (MSW) in \cite{MSW}, where $\sigma_{rec}$ was
treated as a fixed parameter. However, our use of these expressions
is somewhat different.  MSW noted that the existing observations indicate
a value of $\Omega_D\sim 0.6 - 0.7$ and used Eq.~(\ref{disxrec}), with $h=0.7$
to show that this range corresponds to probabilities from $2\%$ to
$12\%$, depending on the values chosen for the galactic scale $M$ and
the spectral index of perturbations $n$.  They concluded that
``anthropic considerations do fairly well as an explanation of a
cosmological constant with $[\Omega_D]$ in the range $0.6-0.7$''.
However, one cannot help but feel disappointed by the somewhat low
values of the probabilities.

Our approach here is that anthropic models should be used as any
other models -- to make testable predictions.  Thus, the goal is
not so much to explain the value of $\Omega_D$ after it is
determined by observations, but to predict that value at a
specified confidence level.  The contour lines in Figs.~2,3 indicate
the $1\sigma$ and $2\sigma$ predictions of the model.  If $M\sim
M_{MW}$ proves to be the relevant mass giving the dominant
contribution to $n_{civ}$, then the currently favoured model with
$\Omega_D\approx .7$ and $h\approx .7$ is virtually excluded by
the anthropic approach at the $2\sigma$ level. Instead, this
approach favours lower values of $h$ and higher values of
$\Omega_D$.

These predictions can be turned around.  If the values
$\Omega_D\leq 0.7$, $h\geq 0.7$ are confirmed by future
measurements, then our model will be ruled out at a 95\%
confidence level, again assuming $M\sim M_{MW}$ and a scale
invariant spectrum. For a tilted spectrum, slightly lower
values of $\Omega_D$ are allowed at the same confidence level. The
observational situation at the time of this writing is far from
being clear. CMB and supernovae measurements yield \cite{Sievers,sn}
$\Omega_D\approx 0.7$ , while the observations of galaxy
clustering give \cite{Bachall} $\Omega_M=0.18\pm 0.8$, and thus
$\Omega_D\approx 0.8$.

\section{The future of the universe}

We finally discuss the anthropic prediction which is not likely to be
tested any time soon.  In all anthropic models, $\rD$ can take both
positive and negative values, so the observed positive dark energy
will eventually start decreasing and will turn negative, and our part
of the universe will recollapse to a big crunch.

To be specific, we shall consider a scalar field model with a very
flat potential.  In the anthropic range (\ref{range}), the potential
can be approximated as a linear function,
\beq
V(\phi)\approx -{V'}_0\phi,
\label{Vlinear}
\eeq
where ${V'}_0$ is a constant and we have set $\phi=0$ at $V=0$.
Once the dark energy dominates, the evolution is described by the
usual slow roll equations
\beq
3H{\dot \phi}={V'}_0,
\label{slow1}
\eeq
\beq
H^2={8\pi \over{3m_p^2}}{V'}_0\phi,
\label{slow2}
\eeq
where $H={\dot a}/a$ and $a(t)$ is the scale factor.
The solution of (\ref{slow1}), (\ref{slow2}) is
\beq
\phi(t)=-\phi_*[1-(t/t_*)]^{2/3},
\eeq
\beq
a(t)=\exp [4\pi m_p^{-2}(\phi_*^2-\phi^2(t))],
\label{at}
\eeq
where $-\phi_*$ is the present value of $\phi$ and
\beq
t_*=8\pi t_D(\phi_*/m_p)^2
\label{t*tD}
\eeq
is the time from the present to the beginning of recollapse.

The slow roll condition (\ref{sr}) implies that $\phi_*\gtrsim m_p$.
As we discussed in Sec. IV, we do not expect this condition to be only
marginally satisfied, and thus $\phi_*\gg m_p$.  Then it follows from
Eqs.~(\ref{t*tD}) and (\ref{at}) that $t_*\gg 8 \pi t_D$ and therefore
we should expect our region of the universe to undergo accelerated
expansion for at least another trillion years before
recollapse.\footnote{This is in contrast with the model of Kallosh and
Linde \cite{kalloshlinde} discussed in Section IV, where the universe
is expected to recollapse within 10-20 billion years.}

The slow roll approximation breaks down at $\phi\sim -m_p$, so the
above equations cannot be used to describe the evolution at $\phi>0$,
where the potential becomes negative.  A general analysis of models
with negative potentials has been given in \cite{Felder}, where it is
shown that at $\phi\gg m_p$ the dynamics becomes dominated by the
kinetic energy of the field, ${\dot\phi}^2\gg |V(\phi)|$.  The corresponding
evolution is described by
\beq
\phi(t)={m_p\over{\sqrt{6\pi}}}\ln(t_c-t) +{\rm const},
\label{crunch1}
\eeq
\beq
a(t)\propto (t_c-t)^{1/3},
\label{crunch2}
\eeq
where $t_c$ is the time of the big crunch.  The linear approximation
(\ref{Vlinear}) for the potential breaks down at sufficiently large
$\phi$, but in this regime the form of the potential is unimportant and
Eqs.~(\ref{crunch1}), (\ref{crunch2}) still apply.

During the dark energy dominated expansion, the ordinary
nonrelativistic matter is diluted by the exponential factor
(\ref{at}).  When the contraction starts, the density of matter begins
to grow as $\rho_M\propto (t_c-t)^{-1}$.  However, the kinetic energy
of the field $\phi$ grows much faster, ${\dot\phi}^2\propto
(t_c-t)^{-2}$, and thus ordinary matter forever remains a subdominant
component of the universe.

\section{Conclusions}

We now summarize the predictions that follow from the anthropic
approach to the CCP's.

(1) In the simplest models where the dark energy density takes discrete values, or where the dark energy density is due to the potential energy of a single scalar field, the dark energy equation of state is predicted to be that of
the vacuum, \beq p_D=w\rho_D, \eeq where $w=-1$ with a very high
accuracy. This distinguishes the anthropic models we discussed
here from other approaches, such as quintessence
\cite{quintessence} or $k$-essence \cite{kessence}.\footnote{For a discussion of multifield models, see \cite{GLV03}. There is a class of two field models where the above prediction does not hold. These are the models where the prior distribution favours a large slope of the potential. In general, the equation of state $p_D=w(z) \rho_D$ depends on the redshift $z$, and increases with time as the dark energy fields pick up kinetic energy, on their way to negative values of the potential. The prediction for $w(z)$ in the general case has been discussed in \cite{kklls,dt}.}

(2) The anthropic predictions for the dark energy density
$\Omega_D$ and for the Hubble parameter $h$ are given in Figs.~2
and 3 of Section V\footnote{These predictions are implicit in the
earlier analysis by Martel, Shapiro and Weinberg \cite{MSW}}. We
show the areas in the $\Omega_d - h$ plane that are excluded at
$1\sigma$ and $2\sigma$ confidence levels. The excluded areas
depend on the assumed galactic mass $M$ and on the spectral index
$n$ of the density fluctuations.  For $M= M_{MW}=10^{12}M_{\odot}$
the currently popular values $\Omega_D=0.7$, $h=0.7$ are
marginally excluded at $2\sigma$ confidence level for a scale
invariant spectrum $n=1$. Lowering the spectral index relaxes the bounds
somewhat. For $h>0.65$ and $n>.95$, the $1\sigma$
prediction is $\Omega_D>0.79$. These anthropic constraints get
weaker when the relevant mass scale $M$ is increased. For example,
with $M=4\times 10^{12}M_{\odot}$ a value as low as
$\Omega_D=0.63$ is still allowed at the $2\sigma$ level for a
scale invariant spectrum. The $1\sigma$ prediction in this case is
$\Omega_D>0.78$ (for $h=0.65$).

(3) Conditions for intelligent life to evolve are expected to arise
mainly in giant galaxies that form (or complete their formation)
at low redshifts, $z_G\lesssim 1$.

(4) The accelerated expansion will eventually stop and our part of the
universe will recollapse, but, at least in the framework of the simplest models, it will take more than a trillion years for this to happen.  Of course, this prediction is not likely to be tested anytime soon. \footnote{Again, this prediction is somewhat different in two field models where the prior distribution favours a large slope of the potential \cite{GLV03}}.

The above predictions apply to models where both CCP's are solved
anthropically. For comparison, we may consider other models. For
instance, it is conceivable that a small value of the cosmological
constant will eventually be explained within the fundamental
theory.  (We note the interesting recent proposal by Dvali,
Gabadadze and Shifman \cite{DGS} in this regard.) Even then, the
coincidence problem will still have to be addressed. One
possibility is that $\rD$ is truly a constant, while the amplitude
of the density fluctuations $\sigma_{rec}$ is a stochastic
variable.  With some mild assumptions about the prior probability
distribution ${\cal P}_*(\sigma_{rec})$, it can be shown
\cite{solutions} that most galaxies are then formed at about the
time of vacuum domination. In this class of models, predictions
(1) and (3) still hold, while the other two predictions no longer
apply.

Another possibility has been recently discussed by Kallosh and Linde
\cite{kalloshlinde}.  They assumed an $M$-theory inspired potential
\beq
V(\phi)=\Lambda(2-cosh\sqrt{2}\phi)
\eeq
with a stochastic variable $\Lambda$.  An interesting property of this
potential is that its curvature is correlated with its height (at
$\phi=0$).  As a result, the universe tends to recollapse within a few
Hubble times after the dark energy comes to dominate.  Assuming that
other contributions to the vacuum energy are somehow cancelled (that
is, that the old CCP is solved by some unspecified mechanism),
Kallosh and Linde argue that the
coincidence $t_D\sim t_I$ is to be expected, where $t_I$ is the time it
takes intelligent life to evolve (they assume it to be $\sim 10^{10}$ yrs).
Predictions (1)-(3) are not applicable to this model.  The model does
predict recollapse of the universe, but the corresponding timescale
($\sim 10^{10}$ yrs) is much shorter than the anthropic prediction (4).

Anthropic arguments are sometimes perceived as handwaving, unpredictive
and unfalsifiable lore, of questionable scientific validity. In our view,
the results presented in this paper should dispel this notion. Here,
we have used the anthropic approach to make several quantitative predictions,
some of which may soon be checked against observations. It should also be
emphasized that, for the particular
case of dark energy, there are at present no alternative theories
explaining both CCP's, or making generic predictions of comparable
accuracy.

The present bound on the equation of state parameter $w$ from the
CMB and supernovae measurements is \cite{Bond} $w<-0.7$, which is
consistent with the anthropic prediction of $w=-1$.  The value of
$w=-1$ is usually associated with a plain cosmological constant.
However, if in addition to this equation of state, observations
confirm some of the other predictions presented above, this may be
taken as an indication that the dark energy is dynamical.  Thus, a
better understanding of structure formation and galactic evolution
may in fact reveal a crucial property of dark energy, with
important implications for particle physics.

\section*{Acknowledgements}

We are grateful to Steven Barr,
David Spergel and Rosanne Di Stefano for useful discussions.
This work was supported by the Templeton Foundation under grant COS
253. J.G. is partially supported by MCYT and FEDER, under
grants FPA2001-3598, FPA2002-00748. A.V. is partially supported by the
National Science Foundation.

\end{document}